\documentclass[twocolumn,aps,prl,showpacs,showkeys,superscriptaddress,notitlepage,longbibliography]{revtex4}
\usepackage[colorlinks=true,urlcolor=blue,citecolor=blue,linkcolor=blue]{hyperref}
\usepackage{graphicx}  
\usepackage{dcolumn}   
\usepackage{bm}        
\usepackage{amssymb}
\usepackage{amsfonts}
\usepackage{doi}
\usepackage{verbatim}
\usepackage{epstopdf}
\usepackage{lineno}
\usepackage{amsmath}
\usepackage{braket}
\usepackage{hyperref}
\DeclareMathOperator{\Tr}{Tr}
\begin{document}
\title{Effects of random domains on the zero Hall plateau in quantum anomalous Hall effect}
\author{Chui-Zhen Chen}
\affiliation{Department of Physics, Hong Kong University of Science and Technology, Clear Water Bay, Hong Kong, China. }
\author{Haiwen Liu}
\affiliation{Center for Advanced Quantum Studies, Department of Physics, Beijing Normal University, Beijing 100875, China. }
\author{X. C. Xie}
\affiliation{International Center for Quantum Materials, School of Physics, Peking University, Beijing 100871, China. }
\affiliation{Collaborative Innovation Center of Quantum Matter, Beijing 100871, China.}
\affiliation{CAS Center for Excellence in Topological Quantum Computation, University of Chinese Academy of Sciences, Beijing 100190, China.}
\begin{abstract}
Recently, a zero Hall conductance plateau with random domains is experimentally observed in quantum anomalous Hall (QAH) effect. We study the effects of random domains on the zero Hall plateau in QAH insulators. We find the structure inversion symmetry determines the scaling property of the zero Hall plateau transition in the QAH systems. In the presence of structure inversion symmetry, the zero Hall plateau state shows a quantum-Hall-type critical point, originating from the two decoupled subsystems with opposite Chern numbers. However, the absence of structure inversion symmetry leads to mixture between these two subsystems, gives rise to a line of critical points, and dramatically changes the scaling behavior. Hereinto, we predict a Berezinskii-Kosterlitz-Thouless-type transition during the Hall conductance plateau switching in the QAH insulators. Our results are instructive for both theoretic understanding of the zero Hall plateau transition and future transport experiments in the QAH insulators.

\end{abstract}

\maketitle


{\emph{Introduction.}}--- Quantum anomalous Hall (QAH) insulator is a new state of  quantum matter and has attracted great interests for both its fundamental and application values \cite{Haldane1988,Onoda2003,Liu2008,Yu2010,Chang2013,Chao-Xing2016,KeHe2018}.
 It possesses a dissipationless chiral edge mode in the bulk gap, giving rise to a quantized Hall conductance.
Initially, the QAH insulator was proposed as a quantum Hall state without external magnetic field \cite{Haldane1988}. 
Later, it was predicated that the QAH effect can be realized in topological insulator (TI) thin film with ferromagnetic (FM) ordering to break the time-reversal symmetry by magnetic doping \cite{Yu2010}. In a recent experiment, the QAH effect was observed by a standard Hall bar measurement in Cr-doped Bi$_x$Sb$_{2-x}$Te$_3$ thin films with a vanishing longitudinal resistance and a quantized Hall resistance plateau \cite{Chang2013}.

\begin{figure}[bht]
\centering
\includegraphics[width=3.3in]{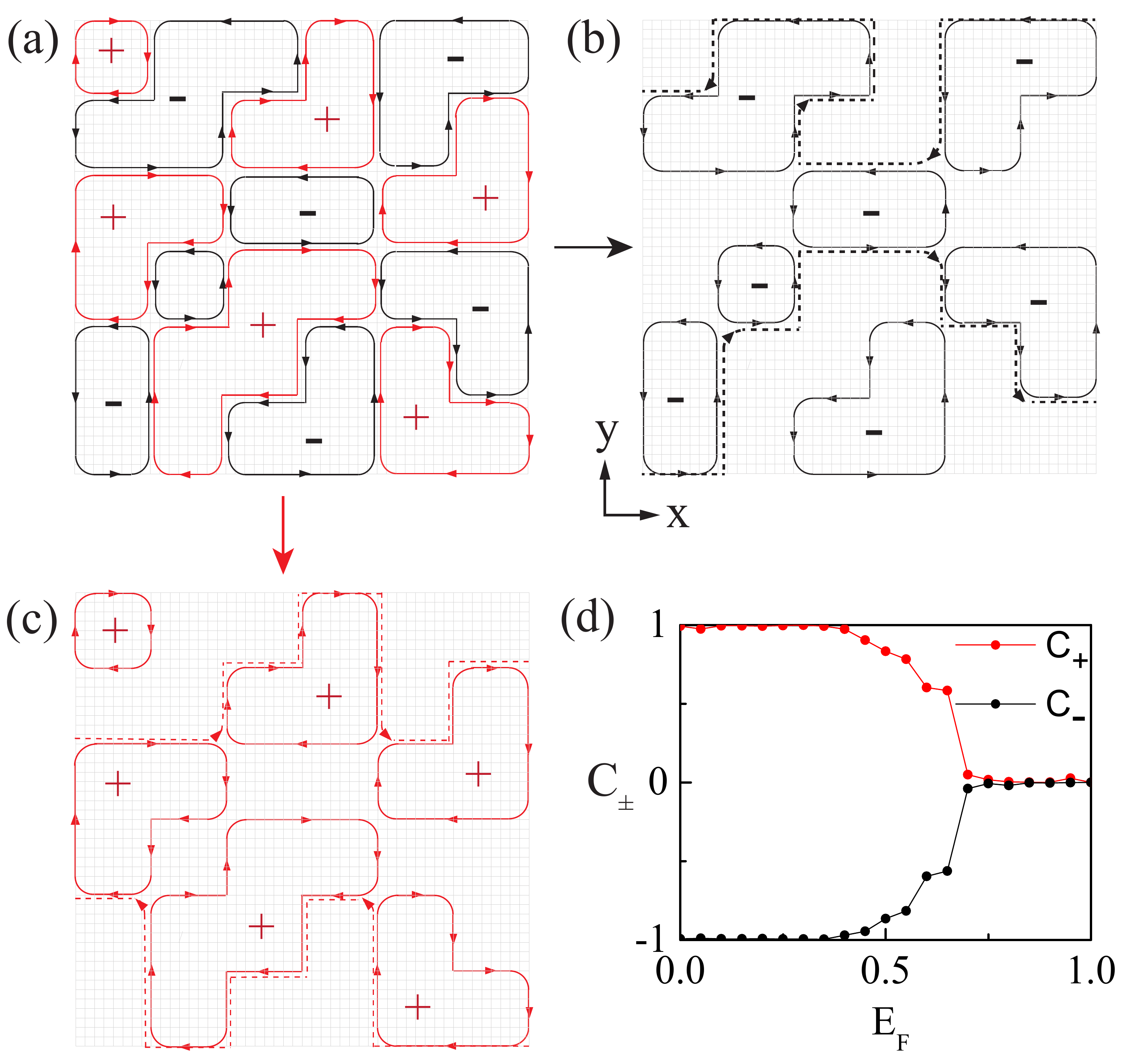}
\caption{(Color online). (a) Schematic plot of zero Hall plateau state on a $48\times48$ magnetic TI lattice with equal population of up ($+$) and down ($-$) magnetic domains in unit of $8\times8$ supercell.
The black and red arrows represent the edge channels propagating on the two domains  species with up (+) and down (-) magnetization.
Due to the structural inversion symmetry, the system can be divided into two subsystems (b) and (c) describing two domain  species, respectively. There is an effective chiral edge channel (dash line) in each of the subsystems.
(d) The two subsystems are topologically equivalent to two Chern insulators with opposite Chern numbers $C_{\pm}=\pm1$.
Therefore, the whole system is topological nontrivial with a quantized spin-Chern number $C_s\equiv(C_+ - C_-)/2$.
Here $E_F$ is the Fermi energy and $C_{\pm}$ are averaged over $20$ random-domain configurations with sample size $96\times96$.
\label{fig1} }
\end{figure}

Plateau transitions between different quantized Hall conductances feature topological properties of QAH effect and have attracted a lot of attention recently \cite{Onoda2003,Wangjing2014,Qiao2016,Checkelsky2014,Wang2015,Kou2015,Chang2016}.
It was experimentally observed that the critical behaviors of plateau transitions between quantized Hall conductance ($\pm e^2/h$) and zero Hall conductance in the QAH effect is qualitatively consistent with  those of quantum Hall effect \cite{Checkelsky2014,Wang2015,Kou2015,Chang2016}.
However, its critical exponent \cite{Wang2015,Kou2015,Chang2016} is deviated from the universal value $\kappa=0.42$ in the quantum Hall effect \cite{Abrahams1981,Wanli2009}.
In these experiments, when the FM ordering induced exchange field $|M_z|$ is greater than the hybridization gap $|m_0|$ due to coupling between the top and bottom surfaces,
the system is a QAH insulator with Chern number $C =M_z/|M_z|$ and the Hall conductance is quantized to be $ C e^2/h$.
Notably, a zero Hall plateau (ZHP) with random magnetic domains shows up during the reversal of magnetization \cite{Lachmane2015,Wuweida2016,Yasuda2017}.
Because the two adjacent magnetic domains have opposite the Chern numbers,
there are two chiral edge modes winding around them in opposite directions, respectively [see black and red lines in Fig.\ref{fig1}(a)].
This is reminiscent of the previous studies of random magnetic field effects on two-dimemsional electron gas \cite{Sheng1995,Xie1997,Xie1998,CWang2015}
and two-channel Chalker-Coddington network model \cite{Chalker1988,Wang1994,Xiong2001}.
Therefore, it is natural to ask if the random magnetic domains can give rise to novel type of phase transitions for the ZHP. Thus, the QAH insulators in FM TI systems provide an ideal platform to study effects of various types of disorder on the scaling properties of plateau transitions.

In this work, we study the effects of random domains on a magnetic TI thin film [see Fig.\ref{fig1}(a)] with disorder.
In the presence of structural inversion symmetry, this system can be divided into two subsystems describing two domain walls species with up and down magnetization, respectively [see Figs.\ref{fig1}(b) and (c)].
It is found that each subsystem with the same  domain species is topological equivalent to a Chern insulator (CI) with Chern number $C_{\pm}=1$ or $-1$ [see Fig.\ref{fig1}(d)].
Under this circumstance, the chiral edge states on the boundaries of the same domain species may tunneling through each other with increasing disorder strength, giving rise a quantum-Hall-type phase transition.
On the other hand, we find that the system undergoes a line of critical points with divergent correlation length $\xi$,
when two domain species are mixed by structural inversion asymmetry (SIA). The SIA is caused by the potential difference between the top and bottom surfaces, and commonly exists in present experimental QAH systems. Moreover, we show that the quantum-Hall-type phase transition or the critical line originate from the robustness of spin-Chern number, which is determined by the existence or absence of the structural inversion symmetry. We predict that without SIA the phase transition between ZHP and the quantized Hall conductance plateau belongs to the Berezinskii-Kosterlitz-Thouless (BKT)-type, which can be verified in the future transport experiments.

{\em Model Hamiltonian.}---We start with a $4\times4$ effective Hamiltonian $H$ of  magnetic doped TI thin film, which can be written as \cite{Yu2010}
\begin{eqnarray}
H \!&=&\! \left(
       \begin{array}{cc}
         h_{+}({\bf k}) &  \\
              & h_{-}({\bf k}) \\
      \end{array}
     \right) + \! H_{SIA} \! \label{Eq1}\\
h_{\pm}({\bf k})\!&=&\!\hbar v_F (k_y\sigma_x \!-\! k_x \sigma_y) \pm m_{\bf k} \sigma_z+\! M_z({\bf r})\sigma_z +\! V_{d}({\bf r})\nonumber
\end{eqnarray}
where $m_{\bf k} = m_0 - m_1 k^2$ are caused by the effective coupling between the top  layer and bottom layer with the momentum ${\bf k}$. The model parameters $m_0$ and $m_1$ are determined by the thickness of  TI films. $v_F$ is the Fermi velocity of the surface states in TI.
Here $\sigma_{x,y,z}$ and $\tau_{x,y,z}$ are the Pauli matrices. 
$M_z({\bf r})$  represents spatial-dependent exchange field in z direction and can simulate the domain effect in magnetic TI \cite{Chen2017}.
The SIA term $H_{SIA}= U_{A}\tau_x  + V_{A}({\bf r})\tau_x $,
where $U_{A}$ and $V_{A}({\bf r})$ measure the uniform and disordered parts of potential difference between the top and bottom surfaces, respectively.
The SIA disorder $V_{A}({\bf r})$ and diagonal disorder $V_{d}({\bf r})$ are independent and they are uniformly distributed in $[-W/2,W/2]$ with the disorder strength $W$.

{\em Quantum-Hall-type transition.}--
In the absence of the SIA term $H_{SIA}$, the Hamiltonian $H$ is block-diagonalized into two subsystems $h_{\pm}$.
In the clean limit, the magnetization is spatially uniform [$M_z({\bf r})=M_z$] and the system is a QAH insulator with total Chern number $C=C_{+}+C_{-}=M_z/|M_z|$ if $|M_z|>|m_0|$.
Here the Chern numbers $C_{\pm}$ of the subsystems $h_{\pm}$ are \cite{Qi2006}
\begin{eqnarray}
\left\{
  \begin{array}{ll}
    C_{+}=1,C_{-}=0 & \hbox{if  $M_z>|m_0|$,} \\
    C_{\pm}=0 & \hbox{if $|M_z|<|m_0|$,} \\
    C_{+}=0,C_{-}=-1 & \hbox{if $M_z<-|m_0|$,}
  \end{array}
\right.
\end{eqnarray}
with $m_{0}<0$ and  $m_{1}>0$.

Next we come to investigate the effects of random domains on the ZHP state in the magnetic TI thin film.
In the simulations, we discrete the model Hamiltonian $H$ on a square lattice (with lattice constant $a=1$)
and set the Fermi velocity $v_F=1$, $m_1=1$, $m_0 = -0.5$, and $|M_z({\bf r})|= 3$.
In Fig.\ref{fig1}(a), the magnetic TI sample is divided into $8\times8$ supercells with the signs of magnetization $M_z({\bf r})$ randomly chosen to be up (+) and down (-) to simulate the random domains.
Due to the structure inversion symmetry, the Hamiltonian $H$ is block diagonalized. Thus the system can be divided into two subsystems to describe two  kinds of random magnetic domains as shown in  Figs.\ref{fig1}(b) and (c), respectively.
Remarkably, we find that the two subsystems are topologically equivalent to two CIs with opposite Chern numbers $C_\pm= \pm1$ as shown in Fig.\ref{fig1}(d).
Therefore, the whole system is topological nontrivial with a quantized spin-Chern number $C_s\equiv(C_+ - C_-)/2$ \cite{Sheng2006,note1}, even though the total Chern number $C=(C_+ + C_-)$ (and thus the Hall conductance) is zero.
Here the  Chern numbers $C_{\pm}$ of the two subsystems are calculated by non-commutative Kubo formula \cite{Prodan2009,Prodan2011}
\begin{eqnarray}
  C_{\pm} &=& 2\pi i \langle \Tr[P_{\pm}[-i[\hat{x},P_{\pm}],-i[\hat{y},P_{\pm}]]]\rangle
\end{eqnarray}
using periodic boundary conditions in both x and y directions, where $\langle...\rangle$ is ensemble-averaged over random configurations and $(\hat{x},\hat{y})$ denotes the position operator.
$P_{\pm}$ is spectral projector onto the positive/negative eigenvalue of $P\tau_zP$ with $P$ the projector onto the occupied states of $H$.
Generally, the Chern numbers $C_{\pm}$ of the two subsystems are quantized, as long as spectrums for both $H$ and $P\tau_zP$ are gapped \cite{Prodan2009,Prodan2011}.
In the presence of structure inversion symmetry, $H$ is block diagonalized in two subsystems $h_\pm$. Thus, $\tau_z$ commutes with $P$ and the spectrum $P\tau_zP$ is gapped and isolated at $\pm1$.
When the SIA term turns on, $\tau_z$ no longer commutes with $P$ and therefore the eigenvalues of $P\tau_zP$ spread between the interval $[-1,1]$. However, the spectrum of $P\tau_zP$ will remain gapped as long as SIA term does not exceed a critical value \cite{Prodan2009,Prodan2011}. 
\begin{figure}[bht]
\centering
\includegraphics[width=3.3in]{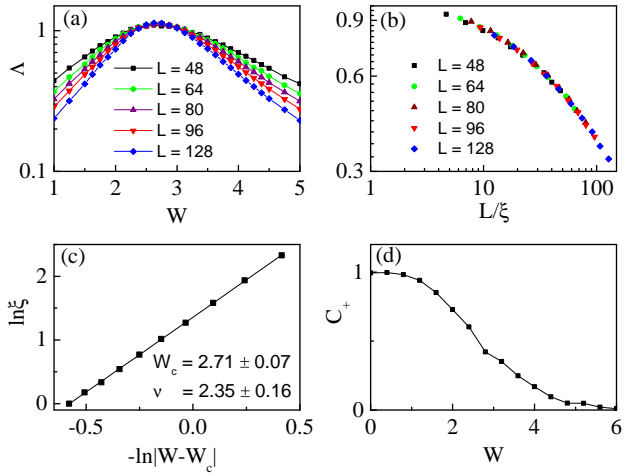}
\caption{(Color online). (a) plots renormalized localization length $\Lambda$ against diagonal disorder strength $W$  (in $V_d$) for $h_+$ with up ($+$) domains as shown in Fig.\ref{fig1}(c).
  (b) shows single parameter scaling of $\Lambda$. All the data on the right side of the critical point collapse to a single curve by a scaling function $\Lambda=f(L/\xi)$ with the correlation length $\xi$. (c) plots $\ln\xi$ against $-\ln|W-W_c|$, which can be fitted with the critical exponent (slope) $\nu=2.35\pm0.16$  and critical disorder strength $W_c=2.71\pm0.07$.
 (d) Chern number $C_{+}$ decreases from one to zeros with increasing disorder strength $W$. The behavior of $C_-$ is similar to $C_+$.
 Fermi energy $E_F = 0.1$ and $C\pm$ are averaged over 40 disordered configurations with sample size $96\times96$.
\label{fig2} }
\end{figure}

To calculate the localization length, we consider a 2D cylinder sample of length $L_x$ and width $L_y=L$ with a periodic boundary condition y direction.
The localization length $\lambda$ is calculated using the transfer matrix method \cite{MacKinnon1981,MacKinnon1983,Kramer1993}. In general, the renormalized localization length $\Lambda\equiv\lambda/L$ increases with $L$ in a metallic
phase, decreases with $L$ in an insulating phase, and is independent of $L$ at the critical point of the phase transition.
For simplicity, we consider the upper block of the system  with random domains as show in Fig.\ref{fig1}(c), which can be described by  the Hamiltonian $h_{+}({\bf k})$ in Eq.1.
In Fig.\ref{fig2}, we find $\Lambda$ decreases with $L$ on both sides of the critical point at diagonal disorder strength $W_c\approx2.7$ [see Fig.\ref{fig2}(a)] while the Chern number $C_+$ decreases by one [see Fig.\ref{fig2}(d)]. This implies a quantum-Hall-type phase transition  between a CI and a normal insulator (NI).
To test the one-parameter scaling theory \cite{MacKinnon1981,MacKinnon1983,Kramer1993}, we show that all the data of $\Lambda$ for $W>W_c$ collapse to a single curve by a scaling function $\Lambda = f(L/\xi)$,
where the correlation length scales by $\xi \propto (W-W_c)^{-\nu}$ with critical exponent $\nu=2.35\pm0.16$ and critical disorder strength $W_c=2.71\pm0.07$.
The critical exponent $\nu$ is in consistence with that of quantum Hall effect \cite{Prange1987,Chalker1988,Huckestein1995,Kramer2005}.  Moreover, because the time-reversal symmetry is restored on average in the presence of random magnetic domains,
lower block of the system $h_{-}({\bf k})$  in Eq.\ref{Eq1} is equivalent to CI with $C_{-}=-1$.
As a consequence, the two sub-blocks of the system with opposite random domain species are topologically equivalent to two CI states with opposite Chern numbers $C_{\pm}=\pm 1$.
We called it spin CI, which has a quantized spin-Chern number $ C_{s}=1$ \cite{Sheng2006,Yang2011,Xu2012}.

\begin{figure}[thb]
\centering
\includegraphics[width=3.3in]{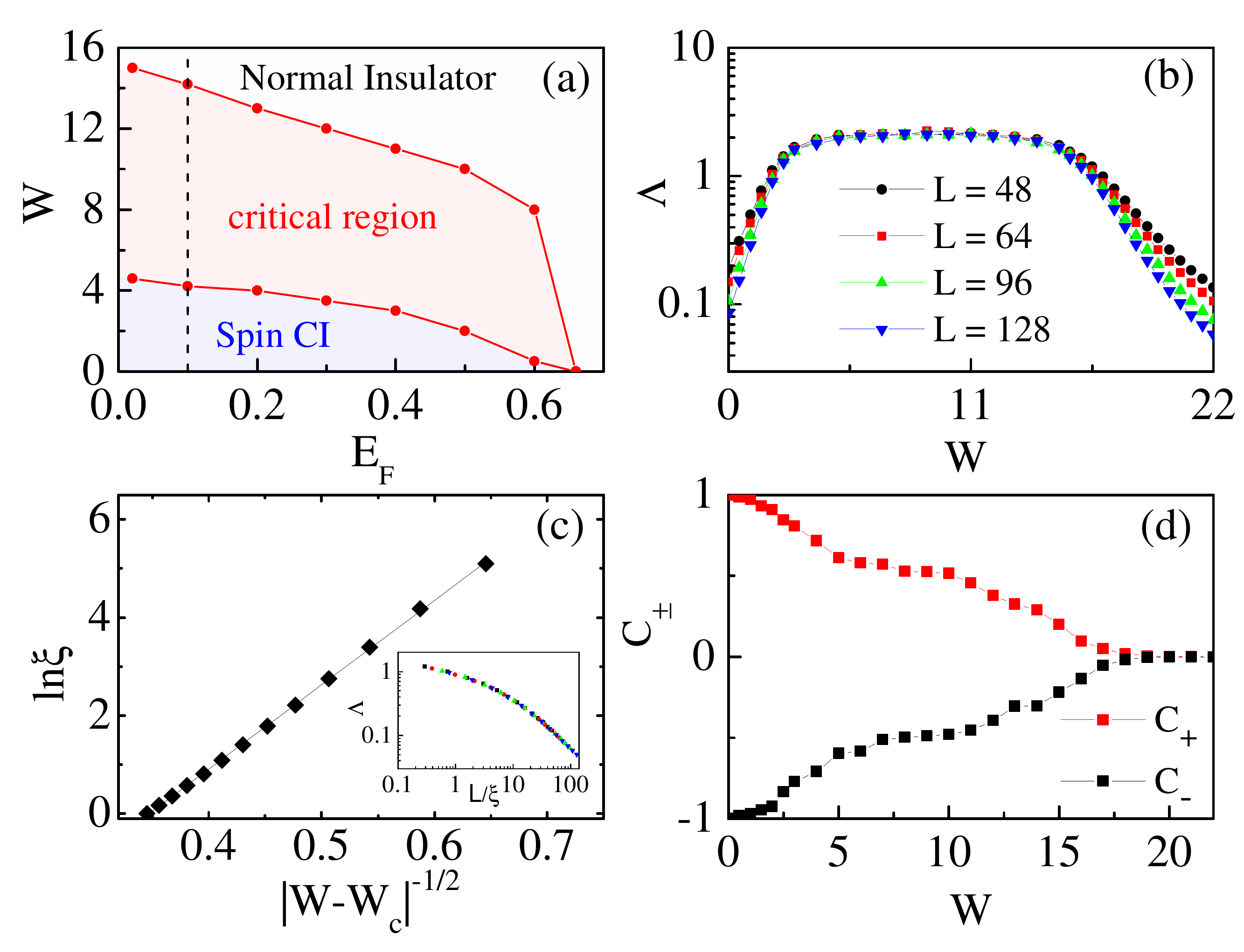}
\caption{(Color online). (a) The phase diagram on the plane of Fermi energy $E_F$ and SIA disorder strength $W$  in $V_A$. The dash line corresponds to the parameter values studied in (b-d). (b) Renormalized localization length $\Lambda$ versus $W$. (c) shows single parameter scaling of $\Lambda$. Logarithmic correlation length $\ln\xi$ is fitted with linearly function of $1/\sqrt{|W-W_c|}$ with $W_c=14.1\pm0.2$, indicating the BKT-type phase transition. The data on the right side of the critical point collapse to a single curve in the inset. (d) plots Chern numbers  $C_{\pm}$ of two-subsystem spaces as a function of SIA disorder strength $W$, correspondingly.
SIA potential difference $U_A=0$ and $C_\pm$ are averaged over $40$ disordered configurations with sample size $96\times96$.
\label{fig3} }
\end{figure}
{\em A line of critical points.}-- Generally, the two sub-block systems $h_\pm$ are coupled by the SIA term $H_{SIA}$ and we shall consider the whole system by Hamiltonian $H$.
When the inversion symmetry is broken by random SIA potential $V_A$, we find a line of critical points with $d\Lambda/dL=0$  between two insulating phases  in Fig.\ref{fig3}(b).
Such a line of critical points coincide with the BKT-type phase transition discovered in  two-dimemsional electron gases with random magnetic field or with random spin-orbit scattering \cite{Xie1997,Xie1998,CWang2015}.
The key  feature  about  the  BKT  transition  is  that the correlation  length $\xi$ diverges as $\xi\propto \exp[\alpha(W-W_c)^{-1/2}]$  on  the  localized  side with critical disorder strength $W_c$ and  parameter $\alpha$ \cite{Xie1997,Xie1998}.
In Fig.\ref{fig3}(c), we find all the data collapse to a single curve (see the inset) and the data for $\ln\xi$ can be fitted with linearly function of $(W-W_c)^{-1/2}$. This supports the transition belonging to the BKT-type. Repeating this procedure at different values of $E_F$ maps out the phase diagram on the $E_F-W$ plane in Fig.\ref{fig3}(a).
In the absent of BIA disorder ($W=0$), the two sub-block systems $h_\pm$ are decoupled and they share the same critical point at $E_F\approx0.66$.
Then $h_\pm$ are coupled by BIA disorder for $W>0$ and the extended states at critical point spread into a critical region.
These extended states go towards band center ($E_F=0$) with increasing $W$, closing the band gap of spin CI at $W\approx4.6$, and they are all localized in strong disorder limit.
Furthermore,  we find that the spin-Chern number $C_s\equiv(C_+-C_-)/2$ is quantized to be one in spin CI phase and then it gradually loses quantization with increasing $W$ in Fig.\ref{fig3}(d).
When the system enters a line of critical points and thus the band gap closes, the spin-Chern number $C_s$ is no longer quantized and it goes zero for NI phase.
The consistency between the phase behaviors obtained from the spin-Chern number and those determined from the localization length demonstrates the reliability of the obtained results.

Next, we provide a phenomenological view to the BKT-type transition discovered above. The BKT transition is a phase transition from the binding to unbinding of vortex-antivortex pairs in the two-dimensional XY model.  It was previously shown that a two-dimensional electron gas in a random magnetic field undergoes a disorder-driven BKT-type metal-insulator transition \cite{Xie1998}. Two different kinds of magnetic domains in the random field system correspond to the vortex and antivortex excitations in the XY model. In the present case, the system possesses random magnetic domains due to random magnetization. These random magnetic domains are divided into the up (+) and down (-) domain species with opposite Chern numbers. This resembles the behaviors of the two-dimensional electron gas in a random magnetic field. Now the BKT transition is driven by the binding and unbinding of the up (+) and down (-) domain species.

\begin{figure}[tbh]
\centering
\includegraphics[width=3.3in]{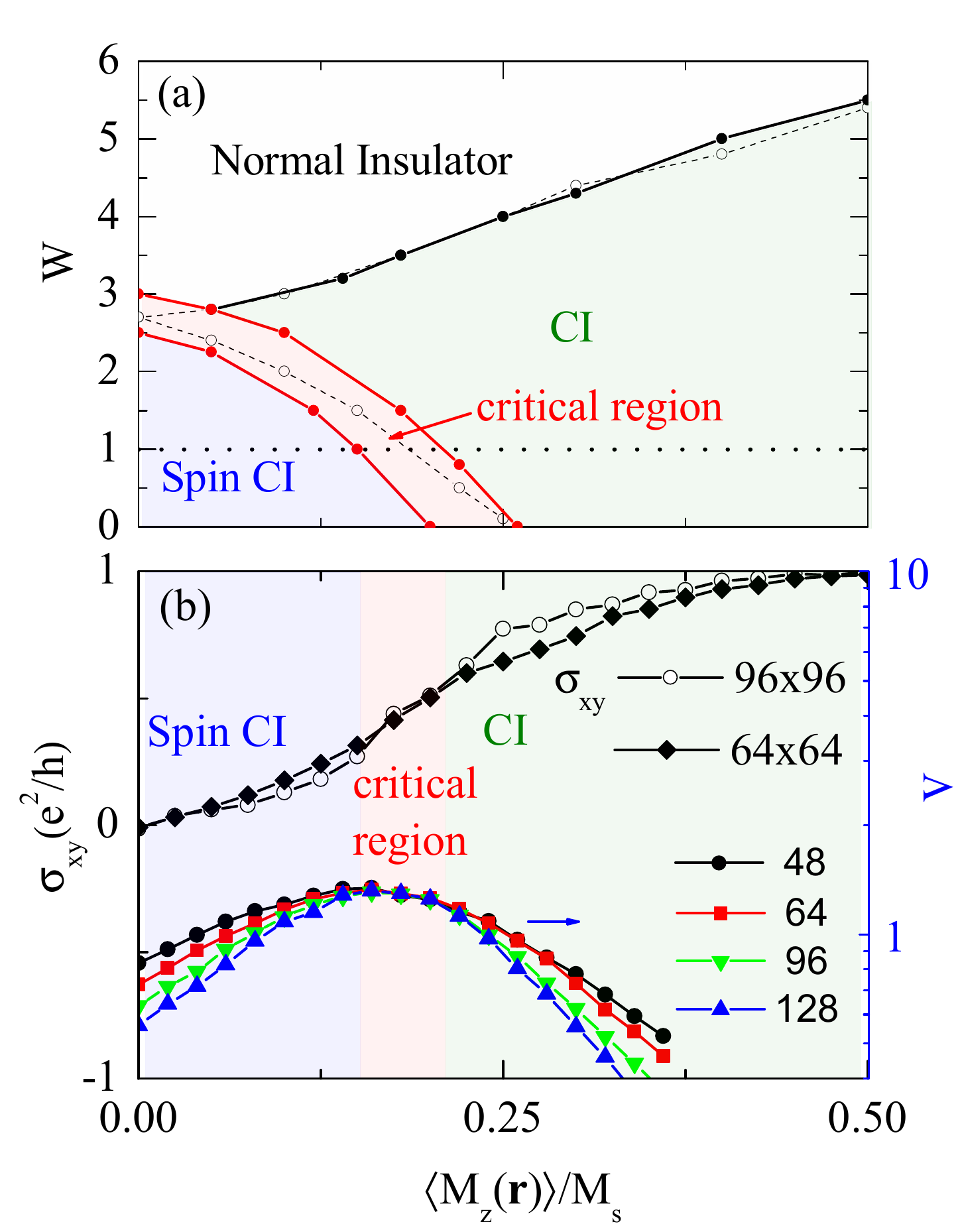}
\caption{(Color online). (a) Phase diagram of magnetic TI with random domains on the  plane of diagonal disorder strength $W$ for $V_d$ and spatial-averaged
normalized magnetization $\langle M_z({\bf r})\rangle/M_s$.
The spin Chern insulator (spin CI), Chern insulator (CI) and  normal insulator (NI) are separated by a critical point or a critical region.
The symbols guided by the solid and dash lines are obtained from
localization length scaling for $U_A=0.05$ and $U_A=0$ with $V_A=0$, respectively.
The dotted line corresponds to the parameter values in (b), which shows disorder-averaged Hall conductance $\sigma_{xy}$ and renormalized localization length $\Lambda$ as a function of $\langle M_z({\bf r})\rangle/M_s$. Other parameters are similar to those in Fig.\ref{fig3}.
\label{fig4} }
\end{figure}
{\em Phase diagram}-- We summarized the main results in the phase diagram on the  plane of  diagonal disorder strength
$W$ (in $V_d$) and spatial averaged magnetization $\langle M_z({\bf r})\rangle$, which is  normalized to saturated magnetization $M_s=3$.
In general, the spin CI is separated from Chern/normal insulator by a single critical point [open dots guided by dash lines in Fig.\ref{fig4}.(a)] or a line of critical point [critical region between red solid lines in Fig.\ref{fig4}.(a)] when the SIA potential difference $U_A=0$ or  $U_A=0.05$, respectively.
For $\langle M_z({\bf r})\rangle=0$, the system is  a spin CI with the equal population of up ($+$) and  down ($-$) domains as we discussed above.
If the structure inversion symmetry is present with $U_A=0$, we find a single critical point between  spin CI and NI phases.
On the other hand, we find a line of critical points before all the states are all localized by the disorder if $U_A=0.05$.
By flipping the magnetization of the domains to up ($+$) direction,
$\langle M_z({\bf r})\rangle$ will gradually increases to $M_s$ until the system is in a single domain state.
During this process, one species of the domain  shrinks while the other expands and the system turns from a spin CI to a CI.
The system shows a critical point or a critical region for $U_A=0$ or  $U_A=0.05$, respectively.

Our numerical results has important implications for transport experiments of the QAH effect.  In Fig.\ref{fig4}(b), we show the Hall conductance $\sigma_{xy}\equiv Ce^2/h$ as a function of spatial averaged magnetization $\langle M_z({\bf r})\rangle$ with $U_A=0.05$ and $W =1$.  We find that a ZHP ($\sigma_{xy}=0$) in the spin CI phase is separated from the quantized Hall conductance $\sigma_{xy}=e^2/h$ by a critical region as indicated by dotted line in the phase diagram in Fig.\ref{fig4}(a). Therefore, we predict that such a phase transition from the ZHP to the quantized Hall conductance plateau belongs to the BKT-type.
We note that the phase transition from the ZHP to the quantized Hall conductance plateau becomes quantum-Hall-type for large disorder $W>3$,
because the spin CI phase is now replaced by the NI phase in Fig.\ref{fig4}(a).

{\em Conclusion}-- In summary, we find that the random domains in magnetic TI has important effects on topological properties of the ZHP state.
The ZHP state with equal population of up and down domains are topologically equivalent to two CIs with opposite Chern numbers.
It is found that the ZHP state goes through a critical point or a critical line in the presence or absence of the structure inversion symmetry, respectively.
Especially for the realistic QAH systems with SIA, we predict a BKT-type phase transition between the ZHP and the quantized Hall plateau.

{\emph{Acknowledgement.}}---  We thank K T Law, Dong-Hui Xu, Emil Prodan, Juntao Song, Ke He and Yang Feng for illuminating discussions.  This work is financially supported by NBRPC (Grants No. 2015CB921102, No. 2017YFA0303301, and
No. 2017YFA0304600), NSFC (Grants No. 11534001, No. 11504008 and No. 11674028), and supported
by the Fundamental Research Funds for the Central Universities.
C.Z.C. thank the support of HKRGC and Croucher Foundation through HKUST3/CRF/13G, 602813, 605512, 16303014 and Croucher Innovation Grant.


\begin{thebibliography}{39}%
\makeatletter
\providecommand \@ifxundefined [1]{%
 \@ifx{#1\undefined}
}%
\providecommand \@ifnum [1]{%
 \ifnum #1\expandafter \@firstoftwo
 \else \expandafter \@secondoftwo
 \fi
}%
\providecommand \@ifx [1]{%
 \ifx #1\expandafter \@firstoftwo
 \else \expandafter \@secondoftwo
 \fi
}%
\providecommand \natexlab [1]{#1}%
\providecommand \enquote  [1]{``#1''}%
\providecommand \bibnamefont  [1]{#1}%
\providecommand \bibfnamefont [1]{#1}%
\providecommand \citenamefont [1]{#1}%
\providecommand \href@noop [0]{\@secondoftwo}%
\providecommand \href [0]{\begingroup \@sanitize@url \@href}%
\providecommand \@href[1]{\@@startlink{#1}\@@href}%
\providecommand \@@href[1]{\endgroup#1\@@endlink}%
\providecommand \@sanitize@url [0]{\catcode `\\12\catcode `\$12\catcode
  `\&12\catcode `\#12\catcode `\^12\catcode `\_12\catcode `\%12\relax}%
\providecommand \@@startlink[1]{}%
\providecommand \@@endlink[0]{}%
\providecommand \url  [0]{\begingroup\@sanitize@url \@url }%
\providecommand \@url [1]{\endgroup\@href {#1}{\urlprefix }}%
\providecommand \urlprefix  [0]{URL }%
\providecommand \Eprint [0]{\href }%
\providecommand \doibase [0]{http://dx.doi.org/}%
\providecommand \selectlanguage [0]{\@gobble}%
\providecommand \bibinfo  [0]{\@secondoftwo}%
\providecommand \bibfield  [0]{\@secondoftwo}%
\providecommand \translation [1]{[#1]}%
\providecommand \BibitemOpen [0]{}%
\providecommand \bibitemStop [0]{}%
\providecommand \bibitemNoStop [0]{.\EOS\space}%
\providecommand \EOS [0]{\spacefactor3000\relax}%
\providecommand \BibitemShut  [1]{\csname bibitem#1\endcsname}%
\let\auto@bib@innerbib\@empty
\bibitem [{\citenamefont {Haldane}(1988)}]{Haldane1988}%
  \BibitemOpen
  \bibfield  {author} {\bibinfo {author} {\bibfnamefont {F.~D.~M.}\
  \bibnamefont {Haldane}},\ }\href {\doibase 10.1103/PhysRevLett.61.2015}
  {\bibfield  {journal} {\bibinfo  {journal} {Phys. Rev. Lett.}\ }\textbf
  {\bibinfo {volume} {61}},\ \bibinfo {pages} {2015} (\bibinfo {year}
  {1988})}\BibitemShut {NoStop}%
\bibitem [{\citenamefont {Onoda}\ and\ \citenamefont
  {Nagaosa}(2003)}]{Onoda2003}%
  \BibitemOpen
  \bibfield  {author} {\bibinfo {author} {\bibfnamefont {M.}~\bibnamefont
  {Onoda}}\ and\ \bibinfo {author} {\bibfnamefont {N.}~\bibnamefont
  {Nagaosa}},\ }\href {\doibase 10.1103/PhysRevLett.90.206601} {\bibfield
  {journal} {\bibinfo  {journal} {Phys. Rev. Lett.}\ }\textbf {\bibinfo
  {volume} {90}},\ \bibinfo {pages} {206601} (\bibinfo {year}
  {2003})}\BibitemShut {NoStop}%
\bibitem [{\citenamefont {Liu}\ \emph {et~al.}(2008)\citenamefont {Liu},
  \citenamefont {Qi}, \citenamefont {Dai}, \citenamefont {Fang},\ and\
  \citenamefont {Zhang}}]{Liu2008}%
  \BibitemOpen
  \bibfield  {author} {\bibinfo {author} {\bibfnamefont {C.-X.}\ \bibnamefont
  {Liu}}, \bibinfo {author} {\bibfnamefont {X.-L.}\ \bibnamefont {Qi}},
  \bibinfo {author} {\bibfnamefont {X.}~\bibnamefont {Dai}}, \bibinfo {author}
  {\bibfnamefont {Z.}~\bibnamefont {Fang}}, \ and\ \bibinfo {author}
  {\bibfnamefont {S.-C.}\ \bibnamefont {Zhang}},\ }\href {\doibase
  10.1103/PhysRevLett.101.146802} {\bibfield  {journal} {\bibinfo  {journal}
  {Phys. Rev. Lett.}\ }\textbf {\bibinfo {volume} {101}},\ \bibinfo {pages}
  {146802} (\bibinfo {year} {2008})}\BibitemShut {NoStop}%
\bibitem [{\citenamefont {Yu}\ \emph {et~al.}(2010)\citenamefont {Yu},
  \citenamefont {Zhang}, \citenamefont {Zhang}, \citenamefont {Zhang},
  \citenamefont {Dai},\ and\ \citenamefont {Fang}}]{Yu2010}%
  \BibitemOpen
  \bibfield  {author} {\bibinfo {author} {\bibfnamefont {R.}~\bibnamefont
  {Yu}}, \bibinfo {author} {\bibfnamefont {W.}~\bibnamefont {Zhang}}, \bibinfo
  {author} {\bibfnamefont {H.-J.}\ \bibnamefont {Zhang}}, \bibinfo {author}
  {\bibfnamefont {S.-C.}\ \bibnamefont {Zhang}}, \bibinfo {author}
  {\bibfnamefont {X.}~\bibnamefont {Dai}}, \ and\ \bibinfo {author}
  {\bibfnamefont {Z.}~\bibnamefont {Fang}},\ }\href {\doibase
  10.1126/science.1187485} {\bibfield  {journal} {\bibinfo  {journal}
  {Science}\ }\textbf {\bibinfo {volume} {329}},\ \bibinfo {pages} {61}
  (\bibinfo {year} {2010})}\BibitemShut {NoStop}%
\bibitem [{\citenamefont {Chang}\ \emph {et~al.}(2013)\citenamefont {Chang},
  \citenamefont {Zhang}, \citenamefont {Feng}, \citenamefont {Shen},
  \citenamefont {Zhang}, \citenamefont {Guo}, \citenamefont {Li}, \citenamefont
  {Ou}, \citenamefont {Wei}, \citenamefont {Wang}, \citenamefont {Ji},
  \citenamefont {Feng}, \citenamefont {Ji}, \citenamefont {Chen}, \citenamefont
  {Jia}, \citenamefont {Dai}, \citenamefont {Fang}, \citenamefont {Zhang},
  \citenamefont {He}, \citenamefont {Wang}, \citenamefont {Lu}, \citenamefont
  {Ma},\ and\ \citenamefont {Xue}}]{Chang2013}%
  \BibitemOpen
  \bibfield  {author} {\bibinfo {author} {\bibfnamefont {C.-Z.}\ \bibnamefont
  {Chang}}, \bibinfo {author} {\bibfnamefont {J.}~\bibnamefont {Zhang}},
  \bibinfo {author} {\bibfnamefont {X.}~\bibnamefont {Feng}}, \bibinfo {author}
  {\bibfnamefont {J.}~\bibnamefont {Shen}}, \bibinfo {author} {\bibfnamefont
  {Z.}~\bibnamefont {Zhang}}, \bibinfo {author} {\bibfnamefont
  {M.}~\bibnamefont {Guo}}, \bibinfo {author} {\bibfnamefont {K.}~\bibnamefont
  {Li}}, \bibinfo {author} {\bibfnamefont {Y.}~\bibnamefont {Ou}}, \bibinfo
  {author} {\bibfnamefont {P.}~\bibnamefont {Wei}}, \bibinfo {author}
  {\bibfnamefont {L.-L.}\ \bibnamefont {Wang}}, \bibinfo {author}
  {\bibfnamefont {Z.-Q.}\ \bibnamefont {Ji}}, \bibinfo {author} {\bibfnamefont
  {Y.}~\bibnamefont {Feng}}, \bibinfo {author} {\bibfnamefont {S.}~\bibnamefont
  {Ji}}, \bibinfo {author} {\bibfnamefont {X.}~\bibnamefont {Chen}}, \bibinfo
  {author} {\bibfnamefont {J.}~\bibnamefont {Jia}}, \bibinfo {author}
  {\bibfnamefont {X.}~\bibnamefont {Dai}}, \bibinfo {author} {\bibfnamefont
  {Z.}~\bibnamefont {Fang}}, \bibinfo {author} {\bibfnamefont {S.-C.}\
  \bibnamefont {Zhang}}, \bibinfo {author} {\bibfnamefont {K.}~\bibnamefont
  {He}}, \bibinfo {author} {\bibfnamefont {Y.}~\bibnamefont {Wang}}, \bibinfo
  {author} {\bibfnamefont {L.}~\bibnamefont {Lu}}, \bibinfo {author}
  {\bibfnamefont {X.-C.}\ \bibnamefont {Ma}}, \ and\ \bibinfo {author}
  {\bibfnamefont {Q.-K.}\ \bibnamefont {Xue}},\ }\href {\doibase
  10.1126/science.1234414} {\bibfield  {journal} {\bibinfo  {journal}
  {Science}\ }\textbf {\bibinfo {volume} {340}},\ \bibinfo {pages} {167}
  (\bibinfo {year} {2013})}\BibitemShut {NoStop}%
\bibitem [{\citenamefont {Liu}\ \emph {et~al.}(2016)\citenamefont {Liu},
  \citenamefont {Zhang},\ and\ \citenamefont {Qi}}]{Chao-Xing2016}%
  \BibitemOpen
  \bibfield  {author} {\bibinfo {author} {\bibfnamefont {C.-X.}\ \bibnamefont
  {Liu}}, \bibinfo {author} {\bibfnamefont {S.-C.}\ \bibnamefont {Zhang}}, \
  and\ \bibinfo {author} {\bibfnamefont {X.-L.}\ \bibnamefont {Qi}},\ }\href
  {\doibase 10.1146/annurev-conmatphys-031115-011417} {\bibfield  {journal}
  {\bibinfo  {journal} {Annual Review of Condensed Matter Physics}\ }\textbf
  {\bibinfo {volume} {7}},\ \bibinfo {pages} {301} (\bibinfo {year}
  {2016})}\BibitemShut {NoStop}%
\bibitem [{\citenamefont {He}\ \emph {et~al.}(2018)\citenamefont {He},
  \citenamefont {Wang},\ and\ \citenamefont {Xue}}]{KeHe2018}%
  \BibitemOpen
  \bibfield  {author} {\bibinfo {author} {\bibfnamefont {K.}~\bibnamefont
  {He}}, \bibinfo {author} {\bibfnamefont {Y.}~\bibnamefont {Wang}}, \ and\
  \bibinfo {author} {\bibfnamefont {Q.-K.}\ \bibnamefont {Xue}},\ }\href
  {\doibase 10.1146/annurev-conmatphys-033117-054144} {\bibfield  {journal}
  {\bibinfo  {journal} {Annual Review of Condensed Matter Physics}\ }\textbf
  {\bibinfo {volume} {9}},\ \bibinfo {pages} {329} (\bibinfo {year}
  {2018})}\BibitemShut {NoStop}%
\bibitem [{\citenamefont {Wang}\ \emph {et~al.}(2014)\citenamefont {Wang},
  \citenamefont {Lian},\ and\ \citenamefont {Zhang}}]{Wangjing2014}%
  \BibitemOpen
  \bibfield  {author} {\bibinfo {author} {\bibfnamefont {J.}~\bibnamefont
  {Wang}}, \bibinfo {author} {\bibfnamefont {B.}~\bibnamefont {Lian}}, \ and\
  \bibinfo {author} {\bibfnamefont {S.-C.}\ \bibnamefont {Zhang}},\ }\href
  {\doibase 10.1103/PhysRevB.89.085106} {\bibfield  {journal} {\bibinfo
  {journal} {Phys. Rev. B}\ }\textbf {\bibinfo {volume} {89}},\ \bibinfo
  {pages} {085106} (\bibinfo {year} {2014})}\BibitemShut {NoStop}%
\bibitem [{\citenamefont {Qiao}\ \emph {et~al.}(2016)\citenamefont {Qiao},
  \citenamefont {Han}, \citenamefont {Zhang}, \citenamefont {Wang},
  \citenamefont {Deng}, \citenamefont {Jiang}, \citenamefont {Yang},
  \citenamefont {Wang},\ and\ \citenamefont {Niu}}]{Qiao2016}%
  \BibitemOpen
  \bibfield  {author} {\bibinfo {author} {\bibfnamefont {Z.}~\bibnamefont
  {Qiao}}, \bibinfo {author} {\bibfnamefont {Y.}~\bibnamefont {Han}}, \bibinfo
  {author} {\bibfnamefont {L.}~\bibnamefont {Zhang}}, \bibinfo {author}
  {\bibfnamefont {K.}~\bibnamefont {Wang}}, \bibinfo {author} {\bibfnamefont
  {X.}~\bibnamefont {Deng}}, \bibinfo {author} {\bibfnamefont {H.}~\bibnamefont
  {Jiang}}, \bibinfo {author} {\bibfnamefont {S.~A.}\ \bibnamefont {Yang}},
  \bibinfo {author} {\bibfnamefont {J.}~\bibnamefont {Wang}}, \ and\ \bibinfo
  {author} {\bibfnamefont {Q.}~\bibnamefont {Niu}},\ }\href {\doibase
  10.1103/PhysRevLett.117.056802} {\bibfield  {journal} {\bibinfo  {journal}
  {Phys. Rev. Lett.}\ }\textbf {\bibinfo {volume} {117}},\ \bibinfo {pages}
  {056802} (\bibinfo {year} {2016})}\BibitemShut {NoStop}%
\bibitem [{\citenamefont {Checkelsky}\ \emph {et~al.}(2014)\citenamefont
  {Checkelsky}, \citenamefont {Yoshimi}, \citenamefont {Tsukazaki},
  \citenamefont {Takahashi}, \citenamefont {Kozuka}, \citenamefont {Falson},
  \citenamefont {Kawasaki},\ and\ \citenamefont {Tokura}}]{Checkelsky2014}%
  \BibitemOpen
  \bibfield  {author} {\bibinfo {author} {\bibfnamefont {J.}~\bibnamefont
  {Checkelsky}}, \bibinfo {author} {\bibfnamefont {R.}~\bibnamefont {Yoshimi}},
  \bibinfo {author} {\bibfnamefont {A.}~\bibnamefont {Tsukazaki}}, \bibinfo
  {author} {\bibfnamefont {K.}~\bibnamefont {Takahashi}}, \bibinfo {author}
  {\bibfnamefont {Y.}~\bibnamefont {Kozuka}}, \bibinfo {author} {\bibfnamefont
  {J.}~\bibnamefont {Falson}}, \bibinfo {author} {\bibfnamefont
  {M.}~\bibnamefont {Kawasaki}}, \ and\ \bibinfo {author} {\bibfnamefont
  {Y.}~\bibnamefont {Tokura}},\ }\href {http://dx.doi.org/10.1038/nphys3053}
  {\bibfield  {journal} {\bibinfo  {journal} {Nature Physics}\ }\textbf
  {\bibinfo {volume} {10}},\ \bibinfo {pages} {731} (\bibinfo {year}
  {2014})}\BibitemShut {NoStop}%
\bibitem [{\citenamefont {Feng}\ \emph {et~al.}(2015)\citenamefont {Feng},
  \citenamefont {Feng}, \citenamefont {Ou}, \citenamefont {Wang}, \citenamefont
  {Liu}, \citenamefont {Zhang}, \citenamefont {Zhao}, \citenamefont {Jiang},
  \citenamefont {Zhang}, \citenamefont {He}, \citenamefont {Ma}, \citenamefont
  {Xue},\ and\ \citenamefont {Wang}}]{Wang2015}%
  \BibitemOpen
  \bibfield  {author} {\bibinfo {author} {\bibfnamefont {Y.}~\bibnamefont
  {Feng}}, \bibinfo {author} {\bibfnamefont {X.}~\bibnamefont {Feng}}, \bibinfo
  {author} {\bibfnamefont {Y.}~\bibnamefont {Ou}}, \bibinfo {author}
  {\bibfnamefont {J.}~\bibnamefont {Wang}}, \bibinfo {author} {\bibfnamefont
  {C.}~\bibnamefont {Liu}}, \bibinfo {author} {\bibfnamefont {L.}~\bibnamefont
  {Zhang}}, \bibinfo {author} {\bibfnamefont {D.}~\bibnamefont {Zhao}},
  \bibinfo {author} {\bibfnamefont {G.}~\bibnamefont {Jiang}}, \bibinfo
  {author} {\bibfnamefont {S.-C.}\ \bibnamefont {Zhang}}, \bibinfo {author}
  {\bibfnamefont {K.}~\bibnamefont {He}}, \bibinfo {author} {\bibfnamefont
  {X.}~\bibnamefont {Ma}}, \bibinfo {author} {\bibfnamefont {Q.-K.}\
  \bibnamefont {Xue}}, \ and\ \bibinfo {author} {\bibfnamefont
  {Y.}~\bibnamefont {Wang}},\ }\href {\doibase 10.1103/PhysRevLett.115.126801}
  {\bibfield  {journal} {\bibinfo  {journal} {Phys. Rev. Lett.}\ }\textbf
  {\bibinfo {volume} {115}},\ \bibinfo {pages} {126801} (\bibinfo {year}
  {2015})}\BibitemShut {NoStop}%
\bibitem [{\citenamefont {Kou}\ \emph {et~al.}(2015)\citenamefont {Kou},
  \citenamefont {Pan}, \citenamefont {Wang}, \citenamefont {Fan}, \citenamefont
  {Choi}, \citenamefont {Lee}, \citenamefont {Nie}, \citenamefont {Murata},
  \citenamefont {Shao}, \citenamefont {Zhang} \emph {et~al.}}]{Kou2015}%
  \BibitemOpen
  \bibfield  {author} {\bibinfo {author} {\bibfnamefont {X.}~\bibnamefont
  {Kou}}, \bibinfo {author} {\bibfnamefont {L.}~\bibnamefont {Pan}}, \bibinfo
  {author} {\bibfnamefont {J.}~\bibnamefont {Wang}}, \bibinfo {author}
  {\bibfnamefont {Y.}~\bibnamefont {Fan}}, \bibinfo {author} {\bibfnamefont
  {E.~S.}\ \bibnamefont {Choi}}, \bibinfo {author} {\bibfnamefont {W.-L.}\
  \bibnamefont {Lee}}, \bibinfo {author} {\bibfnamefont {T.}~\bibnamefont
  {Nie}}, \bibinfo {author} {\bibfnamefont {K.}~\bibnamefont {Murata}},
  \bibinfo {author} {\bibfnamefont {Q.}~\bibnamefont {Shao}}, \bibinfo {author}
  {\bibfnamefont {S.-C.}\ \bibnamefont {Zhang}},  \emph {et~al.},\ }\href
  {http://dx.doi.org/10.1038/ncomms9474} {\bibfield  {journal} {\bibinfo
  {journal} {Nature communications}\ }\textbf {\bibinfo {volume} {6}},\
  \bibinfo {pages} {8474} (\bibinfo {year} {2015})}\BibitemShut {NoStop}%
\bibitem [{\citenamefont {Chang}\ \emph {et~al.}(2016)\citenamefont {Chang},
  \citenamefont {Zhao}, \citenamefont {Li}, \citenamefont {Jain}, \citenamefont
  {Liu}, \citenamefont {Moodera},\ and\ \citenamefont {Chan}}]{Chang2016}%
  \BibitemOpen
  \bibfield  {author} {\bibinfo {author} {\bibfnamefont {C.-Z.}\ \bibnamefont
  {Chang}}, \bibinfo {author} {\bibfnamefont {W.}~\bibnamefont {Zhao}},
  \bibinfo {author} {\bibfnamefont {J.}~\bibnamefont {Li}}, \bibinfo {author}
  {\bibfnamefont {J.~K.}\ \bibnamefont {Jain}}, \bibinfo {author}
  {\bibfnamefont {C.}~\bibnamefont {Liu}}, \bibinfo {author} {\bibfnamefont
  {J.~S.}\ \bibnamefont {Moodera}}, \ and\ \bibinfo {author} {\bibfnamefont
  {M.~H.~W.}\ \bibnamefont {Chan}},\ }\href {\doibase
  10.1103/PhysRevLett.117.126802} {\bibfield  {journal} {\bibinfo  {journal}
  {Phys. Rev. Lett.}\ }\textbf {\bibinfo {volume} {117}},\ \bibinfo {pages}
  {126802} (\bibinfo {year} {2016})}\BibitemShut {NoStop}%
\bibitem [{\citenamefont {Abrahams}\ \emph {et~al.}(1981)\citenamefont
  {Abrahams}, \citenamefont {Anderson}, \citenamefont {Lee},\ and\
  \citenamefont {Ramakrishnan}}]{Abrahams1981}%
  \BibitemOpen
  \bibfield  {author} {\bibinfo {author} {\bibfnamefont {E.}~\bibnamefont
  {Abrahams}}, \bibinfo {author} {\bibfnamefont {P.~W.}\ \bibnamefont
  {Anderson}}, \bibinfo {author} {\bibfnamefont {P.~A.}\ \bibnamefont {Lee}}, \
  and\ \bibinfo {author} {\bibfnamefont {T.~V.}\ \bibnamefont {Ramakrishnan}},\
  }\href {\doibase 10.1103/PhysRevB.24.6783} {\bibfield  {journal} {\bibinfo
  {journal} {Phys. Rev. B}\ }\textbf {\bibinfo {volume} {24}},\ \bibinfo
  {pages} {6783} (\bibinfo {year} {1981})}\BibitemShut {NoStop}%
\bibitem [{\citenamefont {Li}\ \emph {et~al.}(2009)\citenamefont {Li},
  \citenamefont {Vicente}, \citenamefont {Xia}, \citenamefont {Pan},
  \citenamefont {Tsui}, \citenamefont {Pfeiffer},\ and\ \citenamefont
  {West}}]{Wanli2009}%
  \BibitemOpen
  \bibfield  {author} {\bibinfo {author} {\bibfnamefont {W.}~\bibnamefont
  {Li}}, \bibinfo {author} {\bibfnamefont {C.~L.}\ \bibnamefont {Vicente}},
  \bibinfo {author} {\bibfnamefont {J.~S.}\ \bibnamefont {Xia}}, \bibinfo
  {author} {\bibfnamefont {W.}~\bibnamefont {Pan}}, \bibinfo {author}
  {\bibfnamefont {D.~C.}\ \bibnamefont {Tsui}}, \bibinfo {author}
  {\bibfnamefont {L.~N.}\ \bibnamefont {Pfeiffer}}, \ and\ \bibinfo {author}
  {\bibfnamefont {K.~W.}\ \bibnamefont {West}},\ }\href {\doibase
  10.1103/PhysRevLett.102.216801} {\bibfield  {journal} {\bibinfo  {journal}
  {Phys. Rev. Lett.}\ }\textbf {\bibinfo {volume} {102}},\ \bibinfo {pages}
  {216801} (\bibinfo {year} {2009})}\BibitemShut {NoStop}%
\bibitem [{\citenamefont {Lachman}\ \emph {et~al.}(2015)\citenamefont
  {Lachman}, \citenamefont {Young}, \citenamefont {Richardella}, \citenamefont
  {Cuppens}, \citenamefont {Naren}, \citenamefont {Anahory}, \citenamefont
  {Meltzer}, \citenamefont {Kandala}, \citenamefont {Kempinger}, \citenamefont
  {Myasoedov}, \citenamefont {Huber}, \citenamefont {Samarth},\ and\
  \citenamefont {Zeldov}}]{Lachmane2015}%
  \BibitemOpen
  \bibfield  {author} {\bibinfo {author} {\bibfnamefont {E.~O.}\ \bibnamefont
  {Lachman}}, \bibinfo {author} {\bibfnamefont {A.~F.}\ \bibnamefont {Young}},
  \bibinfo {author} {\bibfnamefont {A.}~\bibnamefont {Richardella}}, \bibinfo
  {author} {\bibfnamefont {J.}~\bibnamefont {Cuppens}}, \bibinfo {author}
  {\bibfnamefont {H.~R.}\ \bibnamefont {Naren}}, \bibinfo {author}
  {\bibfnamefont {Y.}~\bibnamefont {Anahory}}, \bibinfo {author} {\bibfnamefont
  {A.~Y.}\ \bibnamefont {Meltzer}}, \bibinfo {author} {\bibfnamefont
  {A.}~\bibnamefont {Kandala}}, \bibinfo {author} {\bibfnamefont
  {S.}~\bibnamefont {Kempinger}}, \bibinfo {author} {\bibfnamefont
  {Y.}~\bibnamefont {Myasoedov}}, \bibinfo {author} {\bibfnamefont {M.~E.}\
  \bibnamefont {Huber}}, \bibinfo {author} {\bibfnamefont {N.}~\bibnamefont
  {Samarth}}, \ and\ \bibinfo {author} {\bibfnamefont {E.}~\bibnamefont
  {Zeldov}},\ }\href {\doibase 10.1126/sciadv.1500740} {\bibfield  {journal}
  {\bibinfo  {journal} {Science Advances}\ }\textbf {\bibinfo {volume} {1}}
  (\bibinfo {year} {2015}),\ 10.1126/sciadv.1500740}\BibitemShut {NoStop}%
\bibitem [{\citenamefont {Wang}\ \emph {et~al.}(2016)\citenamefont {Wang},
  \citenamefont {Chang}, \citenamefont {Moodera},\ and\ \citenamefont
  {Wu}}]{Wuweida2016}%
  \BibitemOpen
  \bibfield  {author} {\bibinfo {author} {\bibfnamefont {W.}~\bibnamefont
  {Wang}}, \bibinfo {author} {\bibfnamefont {C.-Z.}\ \bibnamefont {Chang}},
  \bibinfo {author} {\bibfnamefont {J.~S.}\ \bibnamefont {Moodera}}, \ and\
  \bibinfo {author} {\bibfnamefont {W.}~\bibnamefont {Wu}},\ }\href
  {http://dx.doi.org/10.1038/npjquantmats.2016.23} {\bibfield  {journal}
  {\bibinfo  {journal} {npj Quantum Materials}\ }\textbf {\bibinfo {volume}
  {1}},\ \bibinfo {pages} {16023} (\bibinfo {year} {2016})}\BibitemShut
  {NoStop}%
\bibitem [{\citenamefont {Yasuda}\ \emph {et~al.}(2017)\citenamefont {Yasuda},
  \citenamefont {Mogi}, \citenamefont {Yoshimi}, \citenamefont {Tsukazaki},
  \citenamefont {Takahashi}, \citenamefont {Kawasaki}, \citenamefont {Kagawa},\
  and\ \citenamefont {Tokura}}]{Yasuda2017}%
  \BibitemOpen
  \bibfield  {author} {\bibinfo {author} {\bibfnamefont {K.}~\bibnamefont
  {Yasuda}}, \bibinfo {author} {\bibfnamefont {M.}~\bibnamefont {Mogi}},
  \bibinfo {author} {\bibfnamefont {R.}~\bibnamefont {Yoshimi}}, \bibinfo
  {author} {\bibfnamefont {A.}~\bibnamefont {Tsukazaki}}, \bibinfo {author}
  {\bibfnamefont {K.~S.}\ \bibnamefont {Takahashi}}, \bibinfo {author}
  {\bibfnamefont {M.}~\bibnamefont {Kawasaki}}, \bibinfo {author}
  {\bibfnamefont {F.}~\bibnamefont {Kagawa}}, \ and\ \bibinfo {author}
  {\bibfnamefont {Y.}~\bibnamefont {Tokura}},\ }\href {\doibase
  10.1126/science.aan5991} {\bibfield  {journal} {\bibinfo  {journal}
  {Science}\ }\textbf {\bibinfo {volume} {358}},\ \bibinfo {pages} {1311}
  (\bibinfo {year} {2017})}\BibitemShut {NoStop}%
\bibitem [{\citenamefont {Sheng}\ and\ \citenamefont {Weng}(1995)}]{Sheng1995}%
  \BibitemOpen
  \bibfield  {author} {\bibinfo {author} {\bibfnamefont {D.~N.}\ \bibnamefont
  {Sheng}}\ and\ \bibinfo {author} {\bibfnamefont {Z.~Y.}\ \bibnamefont
  {Weng}},\ }\href {\doibase 10.1103/PhysRevLett.75.2388} {\bibfield  {journal}
  {\bibinfo  {journal} {Phys. Rev. Lett.}\ }\textbf {\bibinfo {volume} {75}},\
  \bibinfo {pages} {2388} (\bibinfo {year} {1995})}\BibitemShut {NoStop}%
\bibitem [{\citenamefont {Liu}\ and\ \citenamefont {Xie}(1997)}]{Xie1997}%
  \BibitemOpen
  \bibfield  {author} {\bibinfo {author} {\bibfnamefont {D.~Z.}\ \bibnamefont
  {Liu}}\ and\ \bibinfo {author} {\bibfnamefont {X.~C.}\ \bibnamefont {Xie}},\
  }\href {\doibase 10.1103/PhysRevB.55.15824} {\bibfield  {journal} {\bibinfo
  {journal} {Phys. Rev. B}\ }\textbf {\bibinfo {volume} {55}},\ \bibinfo
  {pages} {15824} (\bibinfo {year} {1997})}\BibitemShut {NoStop}%
\bibitem [{\citenamefont {Xie}\ \emph {et~al.}(1998)\citenamefont {Xie},
  \citenamefont {Wang},\ and\ \citenamefont {Liu}}]{Xie1998}%
  \BibitemOpen
  \bibfield  {author} {\bibinfo {author} {\bibfnamefont {X.~C.}\ \bibnamefont
  {Xie}}, \bibinfo {author} {\bibfnamefont {X.~R.}\ \bibnamefont {Wang}}, \
  and\ \bibinfo {author} {\bibfnamefont {D.~Z.}\ \bibnamefont {Liu}},\ }\href
  {\doibase 10.1103/PhysRevLett.80.3563} {\bibfield  {journal} {\bibinfo
  {journal} {Phys. Rev. Lett.}\ }\textbf {\bibinfo {volume} {80}},\ \bibinfo
  {pages} {3563} (\bibinfo {year} {1998})}\BibitemShut {NoStop}%
\bibitem [{\citenamefont {Wang}\ \emph {et~al.}(2015)\citenamefont {Wang},
  \citenamefont {Su}, \citenamefont {Avishai}, \citenamefont {Meir},\ and\
  \citenamefont {Wang}}]{CWang2015}%
  \BibitemOpen
  \bibfield  {author} {\bibinfo {author} {\bibfnamefont {C.}~\bibnamefont
  {Wang}}, \bibinfo {author} {\bibfnamefont {Y.}~\bibnamefont {Su}}, \bibinfo
  {author} {\bibfnamefont {Y.}~\bibnamefont {Avishai}}, \bibinfo {author}
  {\bibfnamefont {Y.}~\bibnamefont {Meir}}, \ and\ \bibinfo {author}
  {\bibfnamefont {X.~R.}\ \bibnamefont {Wang}},\ }\href {\doibase
  10.1103/PhysRevLett.114.096803} {\bibfield  {journal} {\bibinfo  {journal}
  {Phys. Rev. Lett.}\ }\textbf {\bibinfo {volume} {114}},\ \bibinfo {pages}
  {096803} (\bibinfo {year} {2015})}\BibitemShut {NoStop}%
\bibitem [{\citenamefont {Chalker}\ and\ \citenamefont
  {Coddington}(1988)}]{Chalker1988}%
  \BibitemOpen
  \bibfield  {author} {\bibinfo {author} {\bibfnamefont {J.~T.}\ \bibnamefont
  {Chalker}}\ and\ \bibinfo {author} {\bibfnamefont {P.~D.}\ \bibnamefont
  {Coddington}},\ }\href {http://stacks.iop.org/0022-3719/21/i=14/a=008}
  {\bibfield  {journal} {\bibinfo  {journal} {Journal of Physics C: Solid State
  Physics}\ }\textbf {\bibinfo {volume} {21}},\ \bibinfo {pages} {2665}
  (\bibinfo {year} {1988})}\BibitemShut {NoStop}%
\bibitem [{\citenamefont {Wang}\ \emph {et~al.}(1994)\citenamefont {Wang},
  \citenamefont {Lee},\ and\ \citenamefont {Wen}}]{Wang1994}%
  \BibitemOpen
  \bibfield  {author} {\bibinfo {author} {\bibfnamefont {Z.}~\bibnamefont
  {Wang}}, \bibinfo {author} {\bibfnamefont {D.-H.}\ \bibnamefont {Lee}}, \
  and\ \bibinfo {author} {\bibfnamefont {X.-G.}\ \bibnamefont {Wen}},\ }\href
  {\doibase 10.1103/PhysRevLett.72.2454} {\bibfield  {journal} {\bibinfo
  {journal} {Phys. Rev. Lett.}\ }\textbf {\bibinfo {volume} {72}},\ \bibinfo
  {pages} {2454} (\bibinfo {year} {1994})}\BibitemShut {NoStop}%
\bibitem [{\citenamefont {Xiong}\ \emph {et~al.}(2001)\citenamefont {Xiong},
  \citenamefont {Wang}, \citenamefont {Niu}, \citenamefont {Tian},\ and\
  \citenamefont {Wang}}]{Xiong2001}%
  \BibitemOpen
  \bibfield  {author} {\bibinfo {author} {\bibfnamefont {G.}~\bibnamefont
  {Xiong}}, \bibinfo {author} {\bibfnamefont {S.-D.}\ \bibnamefont {Wang}},
  \bibinfo {author} {\bibfnamefont {Q.}~\bibnamefont {Niu}}, \bibinfo {author}
  {\bibfnamefont {D.-C.}\ \bibnamefont {Tian}}, \ and\ \bibinfo {author}
  {\bibfnamefont {X.~R.}\ \bibnamefont {Wang}},\ }\href {\doibase
  10.1103/PhysRevLett.87.216802} {\bibfield  {journal} {\bibinfo  {journal}
  {Phys. Rev. Lett.}\ }\textbf {\bibinfo {volume} {87}},\ \bibinfo {pages}
  {216802} (\bibinfo {year} {2001})}\BibitemShut {NoStop}%
\bibitem [{\citenamefont {Chen}\ \emph {et~al.}(2017)\citenamefont {Chen},
  \citenamefont {He}, \citenamefont {Xu},\ and\ \citenamefont
  {Law}}]{Chen2017}%
  \BibitemOpen
  \bibfield  {author} {\bibinfo {author} {\bibfnamefont {C.-Z.}\ \bibnamefont
  {Chen}}, \bibinfo {author} {\bibfnamefont {J.~J.}\ \bibnamefont {He}},
  \bibinfo {author} {\bibfnamefont {D.-H.}\ \bibnamefont {Xu}}, \ and\ \bibinfo
  {author} {\bibfnamefont {K.~T.}\ \bibnamefont {Law}},\ }\href {\doibase
  10.1103/PhysRevB.96.041118} {\bibfield  {journal} {\bibinfo  {journal} {Phys.
  Rev. B}\ }\textbf {\bibinfo {volume} {96}},\ \bibinfo {pages} {041118}
  (\bibinfo {year} {2017})}\BibitemShut {NoStop}%
\bibitem [{\citenamefont {Qi}\ \emph {et~al.}(2006)\citenamefont {Qi},
  \citenamefont {Wu},\ and\ \citenamefont {Zhang}}]{Qi2006}%
  \BibitemOpen
  \bibfield  {author} {\bibinfo {author} {\bibfnamefont {X.-L.}\ \bibnamefont
  {Qi}}, \bibinfo {author} {\bibfnamefont {Y.-S.}\ \bibnamefont {Wu}}, \ and\
  \bibinfo {author} {\bibfnamefont {S.-C.}\ \bibnamefont {Zhang}},\ }\href
  {\doibase 10.1103/PhysRevB.74.085308} {\bibfield  {journal} {\bibinfo
  {journal} {Phys. Rev. B}\ }\textbf {\bibinfo {volume} {74}},\ \bibinfo
  {pages} {085308} (\bibinfo {year} {2006})}\BibitemShut {NoStop}%
\bibitem [{\citenamefont {Sheng}\ \emph {et~al.}(2006)\citenamefont {Sheng},
  \citenamefont {Weng}, \citenamefont {Sheng},\ and\ \citenamefont
  {Haldane}}]{Sheng2006}%
  \BibitemOpen
  \bibfield  {author} {\bibinfo {author} {\bibfnamefont {D.~N.}\ \bibnamefont
  {Sheng}}, \bibinfo {author} {\bibfnamefont {Z.~Y.}\ \bibnamefont {Weng}},
  \bibinfo {author} {\bibfnamefont {L.}~\bibnamefont {Sheng}}, \ and\ \bibinfo
  {author} {\bibfnamefont {F.~D.~M.}\ \bibnamefont {Haldane}},\ }\href
  {\doibase 10.1103/PhysRevLett.97.036808} {\bibfield  {journal} {\bibinfo
  {journal} {Phys. Rev. Lett.}\ }\textbf {\bibinfo {volume} {97}},\ \bibinfo
  {pages} {036808} (\bibinfo {year} {2006})}\BibitemShut {NoStop}%
\bibitem [{not()}]{note1}%
  \BibitemOpen
  \href@noop {} {}\bibinfo {note} {The spin-Chern number is defined in orbital
  space instead of real spin space.}\BibitemShut {Stop}%
\bibitem [{\citenamefont {Prodan}(2009)}]{Prodan2009}%
  \BibitemOpen
  \bibfield  {author} {\bibinfo {author} {\bibfnamefont {E.}~\bibnamefont
  {Prodan}},\ }\href {\doibase 10.1103/PhysRevB.80.125327} {\bibfield
  {journal} {\bibinfo  {journal} {Phys. Rev. B}\ }\textbf {\bibinfo {volume}
  {80}},\ \bibinfo {pages} {125327} (\bibinfo {year} {2009})}\BibitemShut
  {NoStop}%
\bibitem [{\citenamefont {Prodan}(2011)}]{Prodan2011}%
  \BibitemOpen
  \bibfield  {author} {\bibinfo {author} {\bibfnamefont {E.}~\bibnamefont
  {Prodan}},\ }\href {http://stacks.iop.org/1751-8121/44/i=11/a=113001}
  {\bibfield  {journal} {\bibinfo  {journal} {Journal of Physics A:
  Mathematical and Theoretical}\ }\textbf {\bibinfo {volume} {44}},\ \bibinfo
  {pages} {113001} (\bibinfo {year} {2011})}\BibitemShut {NoStop}%
\bibitem [{\citenamefont {MacKinnon}\ and\ \citenamefont
  {Kramer}(1981)}]{MacKinnon1981}%
  \BibitemOpen
  \bibfield  {author} {\bibinfo {author} {\bibfnamefont {A.}~\bibnamefont
  {MacKinnon}}\ and\ \bibinfo {author} {\bibfnamefont {B.}~\bibnamefont
  {Kramer}},\ }\href {\doibase 10.1103/PhysRevLett.47.1546} {\bibfield
  {journal} {\bibinfo  {journal} {Phys. Rev. Lett.}\ }\textbf {\bibinfo
  {volume} {47}},\ \bibinfo {pages} {1546} (\bibinfo {year}
  {1981})}\BibitemShut {NoStop}%
\bibitem [{\citenamefont {MacKinnon}\ and\ \citenamefont
  {Kramer}(1983)}]{MacKinnon1983}%
  \BibitemOpen
  \bibfield  {author} {\bibinfo {author} {\bibfnamefont {A.}~\bibnamefont
  {MacKinnon}}\ and\ \bibinfo {author} {\bibfnamefont {B.}~\bibnamefont
  {Kramer}},\ }\href {\doibase 10.1007/BF01578242} {\bibfield  {journal}
  {\bibinfo  {journal} {Zeitschrift f{\"u}r Physik B Condensed Matter}\
  }\textbf {\bibinfo {volume} {53}},\ \bibinfo {pages} {1} (\bibinfo {year}
  {1983})}\BibitemShut {NoStop}%
\bibitem [{\citenamefont {Kramer}\ and\ \citenamefont
  {MacKinnon}(1993)}]{Kramer1993}%
  \BibitemOpen
  \bibfield  {author} {\bibinfo {author} {\bibfnamefont {B.}~\bibnamefont
  {Kramer}}\ and\ \bibinfo {author} {\bibfnamefont {A.}~\bibnamefont
  {MacKinnon}},\ }\href {http://stacks.iop.org/0034-4885/56/i=12/a=001}
  {\bibfield  {journal} {\bibinfo  {journal} {Reports on Progress in Physics}\
  }\textbf {\bibinfo {volume} {56}},\ \bibinfo {pages} {1469} (\bibinfo {year}
  {1993})}\BibitemShut {NoStop}%
\bibitem [{\citenamefont {Prange}\ and\ \citenamefont
  {Girvin}(1987)}]{Prange1987}%
  \BibitemOpen
  \bibfield  {author} {\bibinfo {author} {\bibfnamefont {R.}~\bibnamefont
  {Prange}}\ and\ \bibinfo {author} {\bibfnamefont {S.}~\bibnamefont
  {Girvin}},\ }\href {https://books.google.com.hk/books?id=Y7XvAAAAMAAJ} {\emph
  {\bibinfo {title} {The Quantum Hall effect}}},\ Graduate texts in
  contemporary physics\ (\bibinfo  {publisher} {Springer-Verlag},\ \bibinfo
  {year} {1987})\BibitemShut {NoStop}%
\bibitem [{\citenamefont {Huckestein}(1995)}]{Huckestein1995}%
  \BibitemOpen
  \bibfield  {author} {\bibinfo {author} {\bibfnamefont {B.}~\bibnamefont
  {Huckestein}},\ }\href {\doibase 10.1103/RevModPhys.67.357} {\bibfield
  {journal} {\bibinfo  {journal} {Rev. Mod. Phys.}\ }\textbf {\bibinfo {volume}
  {67}},\ \bibinfo {pages} {357} (\bibinfo {year} {1995})}\BibitemShut
  {NoStop}%
\bibitem [{\citenamefont {Kramer}\ \emph {et~al.}(2005)\citenamefont {Kramer},
  \citenamefont {Ohtsuki},\ and\ \citenamefont {Kettemann}}]{Kramer2005}%
  \BibitemOpen
  \bibfield  {author} {\bibinfo {author} {\bibfnamefont {B.}~\bibnamefont
  {Kramer}}, \bibinfo {author} {\bibfnamefont {T.}~\bibnamefont {Ohtsuki}}, \
  and\ \bibinfo {author} {\bibfnamefont {S.}~\bibnamefont {Kettemann}},\ }\href
  {\doibase https://doi.org/10.1016/j.physrep.2005.07.001} {\bibfield
  {journal} {\bibinfo  {journal} {Physics Reports}\ }\textbf {\bibinfo {volume}
  {417}},\ \bibinfo {pages} {211 } (\bibinfo {year} {2005})}\BibitemShut
  {NoStop}%
\bibitem [{\citenamefont {Yang}\ \emph {et~al.}(2011)\citenamefont {Yang},
  \citenamefont {Xu}, \citenamefont {Sheng}, \citenamefont {Wang},
  \citenamefont {Xing},\ and\ \citenamefont {Sheng}}]{Yang2011}%
  \BibitemOpen
  \bibfield  {author} {\bibinfo {author} {\bibfnamefont {Y.}~\bibnamefont
  {Yang}}, \bibinfo {author} {\bibfnamefont {Z.}~\bibnamefont {Xu}}, \bibinfo
  {author} {\bibfnamefont {L.}~\bibnamefont {Sheng}}, \bibinfo {author}
  {\bibfnamefont {B.}~\bibnamefont {Wang}}, \bibinfo {author} {\bibfnamefont
  {D.~Y.}\ \bibnamefont {Xing}}, \ and\ \bibinfo {author} {\bibfnamefont
  {D.~N.}\ \bibnamefont {Sheng}},\ }\href {\doibase
  10.1103/PhysRevLett.107.066602} {\bibfield  {journal} {\bibinfo  {journal}
  {Phys. Rev. Lett.}\ }\textbf {\bibinfo {volume} {107}},\ \bibinfo {pages}
  {066602} (\bibinfo {year} {2011})}\BibitemShut {NoStop}%
\bibitem [{\citenamefont {Xu}\ \emph {et~al.}(2012)\citenamefont {Xu},
  \citenamefont {Sheng}, \citenamefont {Xing}, \citenamefont {Prodan},\ and\
  \citenamefont {Sheng}}]{Xu2012}%
  \BibitemOpen
  \bibfield  {author} {\bibinfo {author} {\bibfnamefont {Z.}~\bibnamefont
  {Xu}}, \bibinfo {author} {\bibfnamefont {L.}~\bibnamefont {Sheng}}, \bibinfo
  {author} {\bibfnamefont {D.~Y.}\ \bibnamefont {Xing}}, \bibinfo {author}
  {\bibfnamefont {E.}~\bibnamefont {Prodan}}, \ and\ \bibinfo {author}
  {\bibfnamefont {D.~N.}\ \bibnamefont {Sheng}},\ }\href {\doibase
  10.1103/PhysRevB.85.075115} {\bibfield  {journal} {\bibinfo  {journal} {Phys.
  Rev. B}\ }\textbf {\bibinfo {volume} {85}},\ \bibinfo {pages} {075115}
  (\bibinfo {year} {2012})}\BibitemShut {NoStop}%
\end{thebibliography}
\end{document}